\documentclass[prl, twocolumn, superscriptaddress, reprint, aps]{revtex4-1}

\usepackage{empheq}
\usepackage{amsmath}  
\usepackage{amsfonts}
\usepackage{graphicx}   
\usepackage{hyperref}   
\usepackage{cleveref}

\usepackage{bm}
\usepackage{color}
\newcommand{\be}{\begin{enumerate}}
\newcommand{\ee}{\end{enumerate}}
\newcommand{\bi}{\begin{itemize}}
\newcommand{\ei}{\end{itemize}}

\newcommand{\dif}{\mathrm{d}}

\newcommand{\eq}{\begin{equation}}
\newcommand{\qe}{\end{equation}}
\newcommand{\bal}{\begin{align}}
\newcommand{\lab}{\end{align}}

\usepackage{array}
\usepackage{amssymb}

\begin{document}

\title{Realization of a Topological Phase Transition in a Gyroscopic Lattice}
\author{Noah P. Mitchell}
\email{npmitchell@uchicago.edu}
\thanks{Corresponding author}
\affiliation{James Franck Institute and Department of Physics, University of Chicago, Chicago, IL 60637, USA}
\author{Lisa M. Nash}
\affiliation{James Franck Institute and Department of Physics, University of Chicago, Chicago, IL 60637, USA}
\author{William T. M. Irvine}
\email{wtmirvine@uchicago.edu}
\thanks{Corresponding author}
\affiliation{James Franck Institute and Department of Physics, University of Chicago, Chicago, IL 60637, USA}
\affiliation{Enrico Fermi Institute, The University of Chicago, Chicago, IL 60637, USA}

\begin{abstract}
Topological metamaterials exhibit unusual behaviors at their boundaries, such as unidirectional chiral waves, that are protected by a topological feature of their band structure. The ability to tune such a material through a topological phase transition in real time could enable the use of protected waves for information storage and readout. Here we dynamically tune through a topological phase transition by breaking inversion symmetry in a metamaterial composed of interacting gyroscopes. Through the transition, we track the divergence of the edge modes' localization length and the change in Chern number characterizing the topology of the material's band structure. This work provides a new axis with which to tune the response of mechanical topological metamaterials.
\end{abstract}

\maketitle
A central challenge in physics is understanding and controlling the transport of energy and information.
Topological materials have proven an exceptional tool for this purpose, since topological excitations pass around impurities and defects and are immune to back-scattering at sharp corners~\cite{hasan_colloquium_2010,huber_topological_2016}. 
Furthermore, topological edge modes are robust against weak disorder, such as variations in the pinning energy of each lattice site, in contrast to typical edge waves~\cite{rechtsman_photonic_2013,susstrunk_observation_2015}.

\begin{figure}[h!]
\includegraphics[width=\columnwidth]{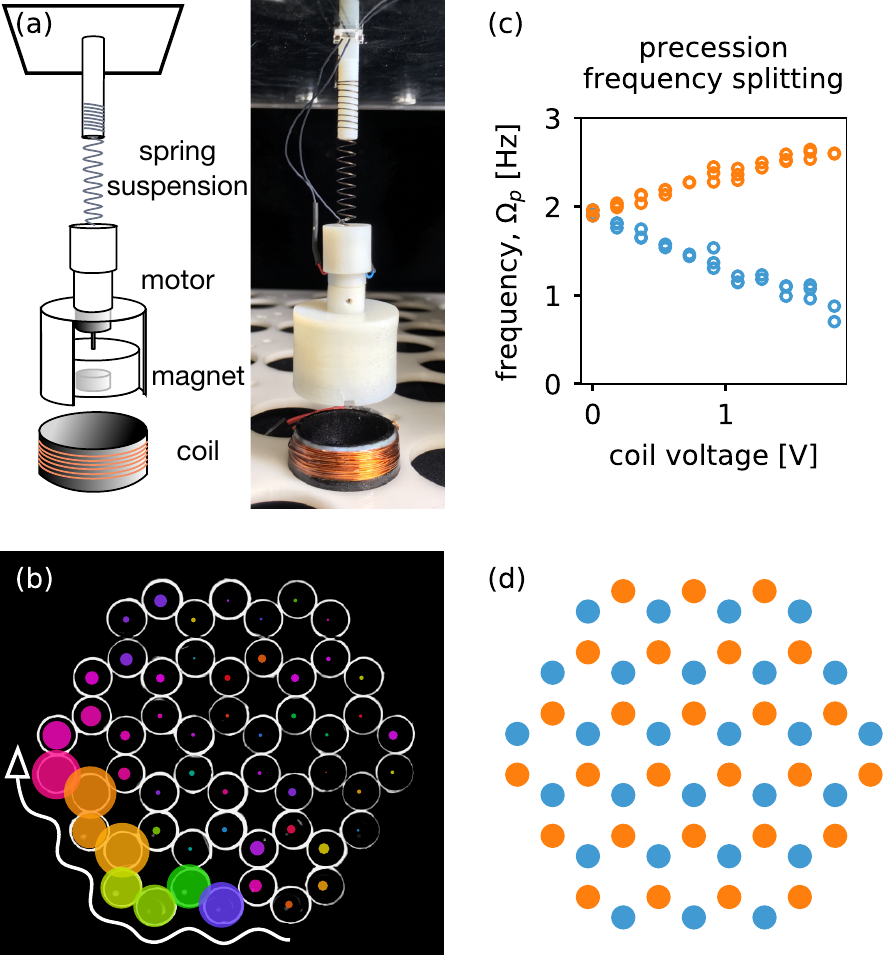}
\caption[]{
\textbf{Modulating the magnetic field at each lattice site tunes a metamaterial of gyroscopes suspended from a plate.}
\textit{(a)} A magnet embedded in each gyro provides an interaction with nearby gyroscopes and with a current-carrying coil.
\textit{(b)} A honeycomb network of interacting gyroscopes supports topologically-protected chiral edge waves. 
Overlaid circles depict the gyroscopes' displacements, colored by the phase of the displacement with respect to the equilibrium positions.
\textit{(c)} The magnetic field from a coil placed below modulates the precession frequency at each site, raising (orange points) or lowering the frequency (blue points) depending on the orientation of the current through the coil.
\textit{(d)} Modulating the precession frequencies in an alternating pattern breaks inversion symmetry of the lattice.
 }
\label{fig1}
\end{figure}

Mechanical topological insulators represent a rapidly-growing class of materials that exhibit topologically-nontrivial phononic band structure~\cite{huber_topological_2016,kane_topological_2013}. 
An important subset of these materials send finite-frequency waves around their perimeter, in a direction determined by the band topology~\cite{prodan_topological_2009,susstrunk_observation_2015,nash_topological_2015,wang_topological_2015,mitchell_amorphous_2016}.
Just as in electronic materials, these edge waves are immune to scattering either into the bulk or in the reverse direction along the edge.
Here we present a method for reversibly passing through a topological phase transition, which allows us to switch chiral edge modes on and off in real time.

\begin{figure*}[ht]
\includegraphics[width=\textwidth]{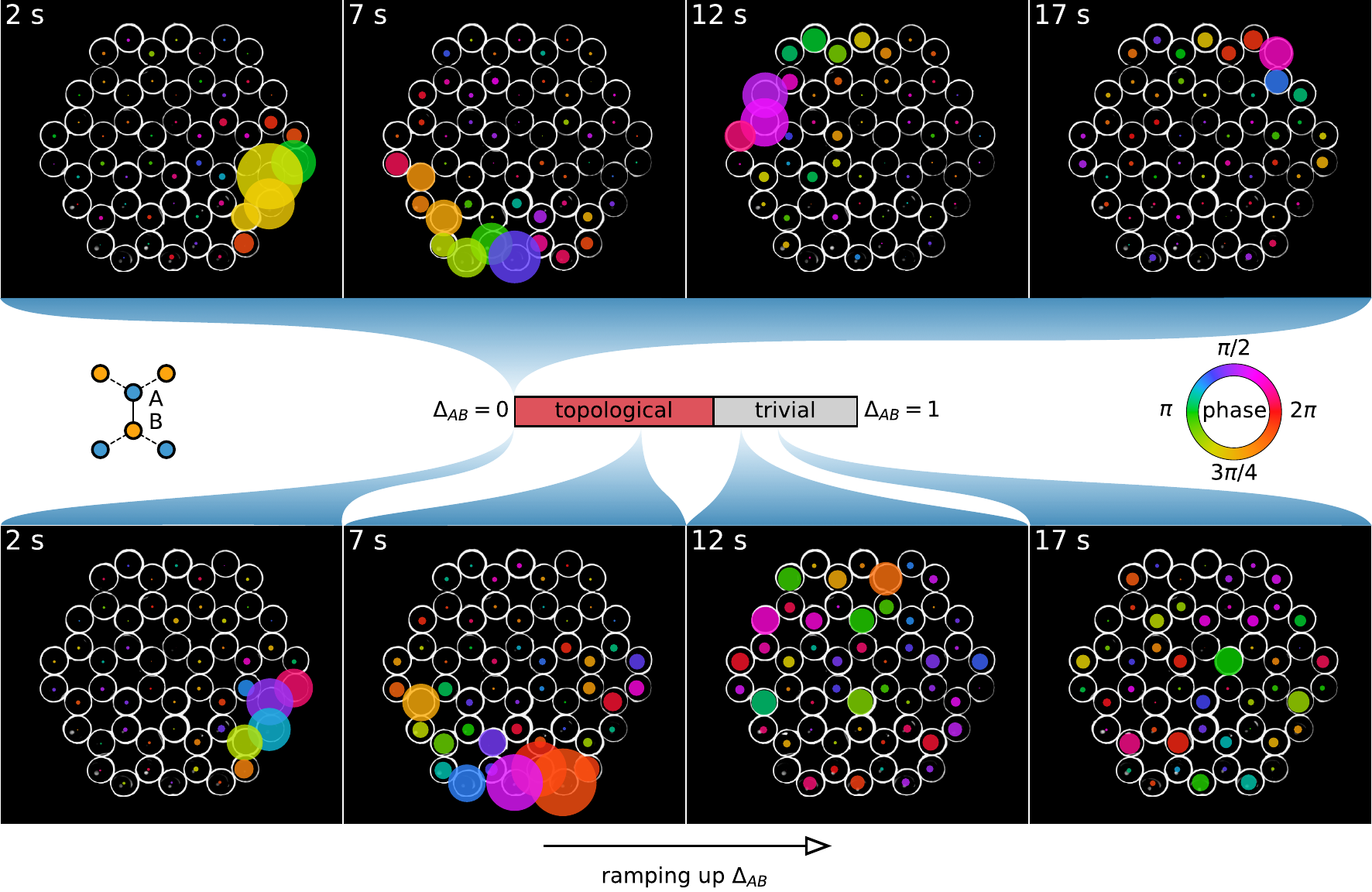}
\caption[]{
\textbf{Dynamically ramping up the inversion symmetry breaking quenches a chiral wave.}
\textit{(top)} For $\Delta_{AB} = 0$, exciting a mode in the gap yields a robust chiral edge wave.
\textit{(bottom)} The same wavepacket is created in the lower panel, but here the inversion symmetry breaking increases over the first 14 seconds of the experiment. 
As $\Delta_{AB}$ passes through the critical value, the mode delocalizes, and no coherent packet persists in the trivial phase. 
The system is viewed from below, through the white coils.
 }
\label{fig2}
\end{figure*}

Our system consists of rapidly-spinning gyroscopes hanging from a plate (Figure~\ref{fig1}). 
If displaced from equilibrium, a single gyroscope will precess: its tip moves in a circular orbit about the equilibrium position as a result of the torques from gravity and the spring suspension.
To induce repulsive interactions between gyroscopes, we place a magnet in each gyroscope with the dipole moments aligned.

A honeycomb lattice of such gyroscopes behaves as a Chern insulator, exhibiting robust chiral edge waves that pass around corners uninhibited.
The phononic spectrum has a band gap, and shaking a boundary site at a frequency in the gap generates a wavepacket that travels clockwise along the edge. 
Figure~\ref{fig1}b shows such a wavepacket as seen from below.

The origin of these chiral edge modes is broken time reversal symmetry, which arises from a combination of lattice structure and spinning components~\cite{nash_topological_2015}.
As in \cite{nash_topological_2015,mitchell_amorphous_2016}, an effective time reversal operation both reverses time ($t\rightarrow-t$) and reflects one component of each gyroscope's displacement ($\psi \rightarrow \psi^*$, where $\psi = \delta x + i \delta y$ is the displacement of a gyroscope).
Breaking effective time reversal symmetry opens a gap at the Dirac points of the phononic dispersion in a way that endows each band with a nonzero Chern number~\cite{haldane_model_1988}. 

An alternative mechanism for opening a gap, however, is to make sites in the unit cell inequivalent.
This process breaks inversion symmetry in the honeycomb lattice: the system is no longer invariant under exchange of the two sites in the unit cell.
A gap opened by this mechanism is topologically trivial and does not lead to protected edge modes.

If both symmetries are broken, then their relative strength should determine whether the system is topological or trivial, enabling us to tune the system through a phase transition.
This would be analogous to a known transition in the Haldane model~\cite{haldane_model_1988}, in which broken inversion symmetry competes with the broken time reversal symmetry.

A simple way to break inversion symmetry is to detune the precession frequencies of neighboring gyroscopes, pairwise throughout the system.
To do so, we apply a local magnetic field at each site by introducing a coil beneath each gyroscope.
For small displacements, the coil's magnetic field provides a force which is parallel or antiparallel to gravity, raising or lowering the its precession frequency: $\Omega_p \rightarrow \Omega_p^{A,B}= (1 \pm \Delta_{AB}) \Omega_p^0$ (Figure~\ref{fig1}c).
We then assemble a honeycomb lattice of gyroscopes with alternating coil orientations at each site (Figure~\ref{fig1}d).
To reduce noise in the precession frequencies, we synchronize all spinning speeds by sending pulse-width-modulated signals to the motors.

We excite a wavepacket in this system, again by shaking a site at the boundary, and simultaneously ramp up the current through the coils. 
As we pass through a critical current --- corresponding to a critical inversion symmetry breaking strength --- the excitation delocalizes:
the coherent, topologically-protected edge mode transforms into bulk modes, suggesting the presence of a topological phase transition (see bottom panel in Figure~\ref{fig2}).


\begin{figure}[ht]
\includegraphics[width=\columnwidth]{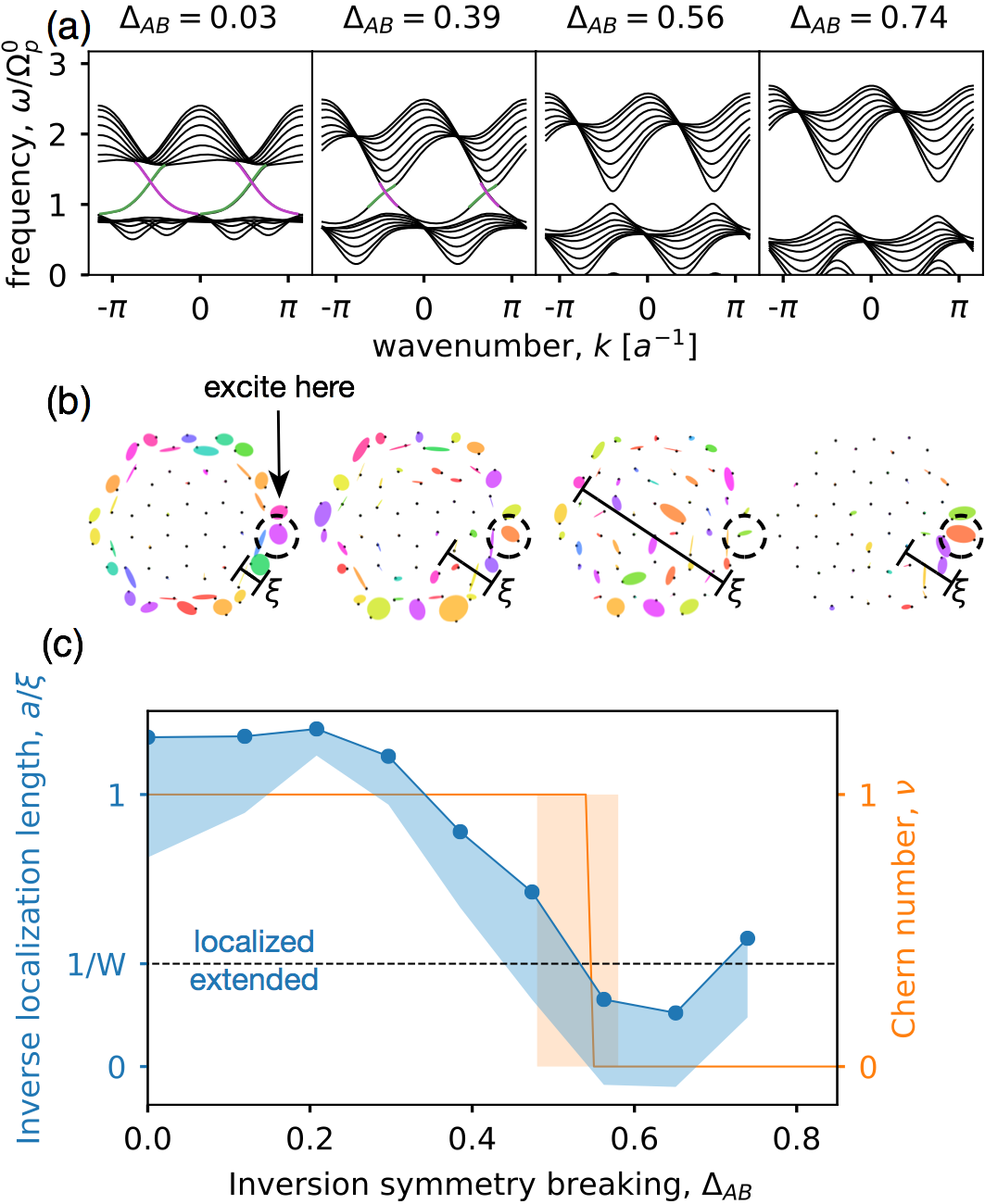}
\caption[]{
\textbf{Inversion symmetry breaking drives the topological transition to a trivial insulator, with a divergence in localization length at the transition.}
\textit{(a)} Band structure for a periodic supercell shows the gap close and reopen; increasing the frequency splitting, $\Delta_{AB}$, eliminates the chiral edge modes localized to the top (purple) and bottom (green) of the supercell.
\textit{(b)} Experimental measurement of the edge states near the center of the gap show the divergence of the localization length. 
At large $\Delta_{AB}$, exciting the system in the trivial band gap leads to a  weak, localized response.
\textit{(c)} When the most localized state (connected blue dots) becomes extended such that $\xi \sim W$ (black dashed line), where $W$ is the distance from the center to the system's boundary, the Chern number of the lower band (orange line) changes from $+1$ to zero. The orange band represents uncertainty in the transition point arising from uncertainty in experimental parameters.
All states in the range $[0.9\, \Omega_p^0, 1.1\,\Omega_p^0]$ are included in the blue band, and $a$ is the lattice spacing.
 }
\label{fig3}
\end{figure}

\begin{figure}[ht]
\includegraphics[width=\columnwidth]{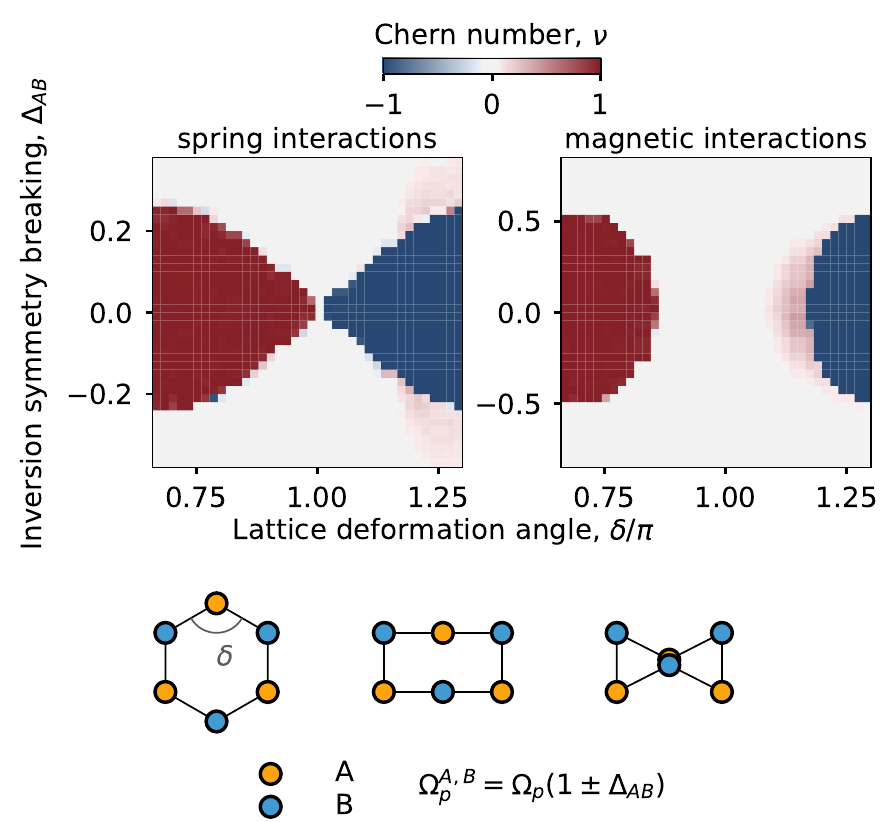}
\caption[]{
\textbf{Our experiment is one slice of a larger phase diagram.}
Tuning the lattice geometry to a bricklayer lattice restores effective time reversal symmetry, and continuing to deform into a bowtie inverts the sign of the symmetry-breaking term.
The left panel shows the topological phase diagram for the case of spring couplings (no restoring force in the transverse direction, to first order), with $\Omega_p^0 =\Omega^{+}=\Omega^{-}$. 
The right panel shows the Chern number of the lower band for the case of magnetic interactions, with $\Omega_k/ \Omega_p^0 = 0.67$, as in the experiments. 
The two are similar, though the topologically nontrivial phases (red and blue) do not meet at a point in the case of magnetic interactions.
 }
\label{fig4}
\end{figure}

To study the transition in more detail, we compute the band structure of magnetic gyroscopes with varying inversion symmetry breaking.
As the precession frequency splitting is increased, the band gap closes and reopens (Figure~\ref{fig3}a).
Beyond the critical value, no edge states connect the two bands: the system is a trivial insulator.

Measuring the localization length of edge states in our experiment enables a direct comparison against our model. 
As the gap narrows, the localization length of the most confined edge state broadens until it is comparable to the system size. 
By shaking the system at a frequency that slowly sweeps through the gap and tracking the gyroscopes' displacements, we obtain the eigenstates of the system (the eigenvectors of the Fourier transform).
Figure~\ref{fig3}b and c show the results of this measurement, considering only the most localized states near the center of the gap.
As the localization length of the most localized state (blue curve in Figure~\ref{fig3}c) grows to the scale of the system size, $W$, the bands touch and reopen without chiral edge modes. 
This feature confirms the likely presence of a topological phase transition.

To predict the topological phase transition theoretically, we compute the Chern number of the system's bands, which is encoded in the spectrum of the dynamical matrix. 
To linear order, the displacement of a fast-spinning gyroscope, $\psi \equiv \delta x + i \delta y$, obeys Newton's second law as
\begin{equation}
	\label{eom}
    \begin{split}
	i \frac{d \psi_p}{d t} = 
    \Omega_p \psi + \frac{1}{2} \sum_q 
    \bigg[ & \left(\Omega_{pp}^{+} \psi_p + \Omega_{pq}^{+} \psi_q\right)   \\
    & + e^{2 i \theta_{pq}}\left(\Omega_{pp}^{-} \psi^*_p 
+ \Omega_{pq}^{-} \psi^*_q\right) \bigg],
	\end{split}
\end{equation}
where the sum is over nearby gyroscopes, $\Omega^{\pm}_{pq} \equiv - \frac{\ell^2}{I\omega}\left( \partial F_{p\parallel} /\partial x_{q \parallel} \pm \partial F_{p\perp}/\partial x_{q\perp} \right)$
 is the characteristic interaction frequency between gyroscopes $p$ and $q$, 
$\Omega_p \equiv (mg + F^{\textrm{suspension}} + F^{\textrm{coil}}_z) \ell/I\omega = (1+\Delta_{AB}) \Omega_p^0$ is the precession frequency in the absence of other gyroscopes, 
and $\theta_{pq}$ is the angle of the bond connecting gyroscope $p$ to gyroscope $q$, taken with respect to a fixed global axis.
The interaction strengths, $\Omega^{\pm}_{pq}$, scale with the quantity $\Omega_k \equiv \ell^2 k_m/I\omega$, where $k_m$ is the effective spring constant for the magnetic interaction, and $\Omega^{\pm}_{pq}$ depend nonlinearly on the lattice spacing. 
As equation~\ref{eom} resembles the Schr\"{o}dinger equation, we write the equation of motion for the entire system as
\begin{equation}
i \frac{d\vec{\psi}}{dt} = D \vec{\psi}.
\end{equation}
The precession frequency plays the role of the on-site potential, so that the coil's magnetic field detunes the diagonal terms of the dynamical matrix, $D$.

A nonzero Chern number signals the existence of topologically-protected chiral edge modes.
Computing the Chern number of the magnetic system as
\begin{equation}
C_j \dif x \wedge \dif y = \frac{i}{2\pi} \int \dif^2 k \,\textrm{Tr} \left(\dif P_j \wedge P_j \dif P_j \right),
\end{equation}
where the projector $P_j \equiv | u_j \rangle \langle u_j |$ maps states in band $j$ to themselves and maps other states to zero, 
we see the Chern number of the lower band change from 1 to 0 when the localization length reaches the system size, $\xi \sim W$ (Figure~\ref{fig3}c).

Our approach enables us to tune through a phase transition dynamically by pitting inversion symmetry against time reversal symmetry breaking, adding a new axis of versatility for topological mechanical metamaterials.
We illustrate this by computing a larger phase diagram for the gyroscopic system in which time reversal symmetry breaking and inversion symmetry breaking are both varied.
To explore their interplay, we combine the transition observed here with another topological phase transition discussed in~\cite{nash_topological_2015}, which exploits the dependence of the band topology on the geometry of the lattice. 
By globally deforming the honeycomb lattice through a brick-layer lattice, the Chern number of the lower band transitions from 1 to 0 to -1 (Figure~\ref{fig4}). 
The transition occurs when bond angles in the network are precisely multiples of $\pi/2$, at which point effective time reversal symmetry is restored and the gap closes.
Continuing the deformation into a bowtie configuration inverts the sign of the symmetry-breaking term, reopening the gap, but each band's Chern number flips sign.

In Figure~\ref{fig4}, we allow inversion symmetry breaking and lattice deformation to compete, giving rise to systems with clockwise edge modes (red), counterclockwise edge modes (blue), and no chiral edge modes (white). 
When the time reversal symmetry breaking is weakened ($\delta \rightarrow \pi$), the required $|\Delta_{AB}|$ to drive the system to a trivial insulator diminishes.
This phase diagram highlights the similarity of the gyroscopic system to the Haldane model~\cite{haldane_model_1988}.
In particular, the case with spring interactions, in which $\Omega^{+} = \Omega^{-}$, is similar to Figure 2 in reference~\cite{haldane_model_1988}.
This can be understood by taking the limit in which the precession frequency is much faster than the interaction strength ($\Omega_p^0 \gg \Omega_k$): the spring-coupled gyroscope system maps to the Haldane model~\cite{nash_topological_2015}.

The phase diagram for magnetic interactions, while similar, possesses an area of trivial insulator between the topologically nontrivial phases.
Unlike spring-like potentials, magnets exhibit an anti-restoring response to perpendicular displacements ($\Omega^{+} \ne \Omega^{-}$). 
These interactions can introduce an indirect band gap that closes before the Dirac points touch. 
The result is a trivial insulator phase of finite extent separating the phases with $\nu = +1$ and $\nu=-1$ for the lower band (right panel in Figure~\ref{fig4}).
The extent of the separation depends on the interaction strength $\Omega^{\pm}/\Omega_p^0$ and the spacing between gyroscopes relative to their pendulum length.

In conclusion, we have demonstrated a non-destructive mechanism for dynamically tuning a mechanical Chern insulator through a topological phase transition. 
We characterized the transition by measuring the delocalization of edge modes in gap and the corresponding change in Chern number to zero at the transition, and we established this 1D transition's context within a 2D phase space for mechanical gyroscopic Chern insulators.
This design enables a mechanism for constructing topological gates able to store and read out information in chiral modes~\cite{bilal_bistable_2017}.

\textbf{Note:} We would like to point out a parallel study of a different topological phase transition in a photonic system, in which time is replaced by propagation through a cleverly-designed waveguide structure~\cite{guglielmon_transition_2017}.


\end{document}